\documentstyle[12pt]{article}
\def\slash#1{#1 \hskip -0.5em / }
\global\arraycolsep=2pt %reduces the separation in eqnarrays
\input{epsf}
\begin{document}
%--------------------------------
%---------------- CERN Titlepage <---------------------------
\thispagestyle{empty}
\begin{titlepage}

\begin{flushright}
CERN-TH.7449/94
\end{flushright}

\vspace{0.3cm}

\begin{center}
\Large\bf The $1/m_Q$ Expansion in QCD:  \\
       Introduction and Overview
\end{center}

\vspace{0.8cm}

\begin{center}
Thomas Mannel  \\
{\sl Theory Division, CERN, CH-1211 Geneva 23, Switzerland}
\end{center}

\vspace{0.8cm}

\begin{abstract}
\noindent
A mini-review of the heavy mass expansion in QCD is given. We
focus on exclusive semileptonic decays and some topics of recent interest
in inclusive decays of heavy hadrons.
\end{abstract}
\vfill
\centerline{Contribution to the Workshop {\it QCD 94},
                 Montpellier, 7-13 July 1994}
\bigskip\bigskip
\vfill
\noindent
CERN-TH.7449/94 \\
September 1994
\end{titlepage}

%---------------- END CERN Titlepage <---------------------------
\section{Introduction}
\noindent
In the past five years considerable progress has been made towards
a QCD-based and model independent description of hadrons containing
a heavy-flavour quark. This progress has been achieved by the use of
the heavy mass limit for the heavy quark, in which it is replaced by
a static source of a colour field. This limit of QCD leads to a
well-defined field theory, the so-called Heavy Quark Effective Theory (HQET).

The infinite mass limit from QCD has been used for some time for various
purposes \cite{shifman}, but the main observation, which triggered the
enormous development of this field, is that in the infinite mass limit
two additional symmetries appear, which are not present in full QCD
\cite{isgur/wise}. Since then the implications of the heavy mass limit
and HQET have been extensively studied in innumerable publications, and
the development of the field is documented in more or less extensive
reviews \cite{reviews}.

The two additional symmetries of the heavy mass limit have important
phenomenological applications; they lead to model-independent
relations between
form factors describing e.g. exclusive weak decays. The origin of
the new symmetries
is quite simple. The first symmetry is a heavy-flavour symmetry, which is
due to the fact that the interaction of the quarks with
the gluons is flavour-blind and in the heavy mass limit all heavy quarks
act as a static source of colour. Formally this corresponds to an $SU(2)$
symmetry relating $b$ and $c$ quarks moving with the same velocity.
The second symmetry is the spin symmetry of the heavy quark.
The interaction of the heavy quark spin with
the ``chromomagnetic'' field is inversely proportional to the heavy mass
and hence vanishes in the infinite mass limit. As a consequence, the
rotations for the heavy quark spin become an $SU(2)$ symmetry, which
holds for a fixed velocity of the heavy quark.

Corrections to the limit $m_Q \to \infty$ may be studied systematically
in the framework of HQET. The corrections
are given as power series expansions in two small parameters. The
first small parameter is the strong coupling constant,
taken at the scale of the heavy quark $\alpha_s (m_Q)$. This type
of correction may be calculated systematically using perturbation
theory in HQET.
The second type of correction is characterized by the small parameter
$\bar\Lambda / m_Q$, where $\bar\Lambda$ is a scale of
the light QCD degrees of freedom, e.g. $\bar\Lambda \sim m_{hadron}-m_Q$.
In the effective theory approach this type of corrections enter
through operators of higher dimension,
the matrix elements of which have to be parametrized in general
by additional form factors.

More or less parallel to the development of heavy quark symmetry and
HQET, which is well suited for exclusive decays, the heavy mass expansion
has been applied also to inclusive decays \cite{russnl}-\cite{FLS93}.
The heavy quark mass
sets a scale that is large compared to $\Lambda_{QCD}$,
and one may use a similar
setup as in deep inelastic scattering for the inclusive decays; in
particular, the operator product expansion for the inclusive decays
yields an expansion in inverse powers of the heavy quark mass. In this
way one may not only study total rates, but also differential distributions
such as the lepton energy spectra in inclusive semileptonic decays.

This mini-review is intended to set up the stage for some of the contributions
to this conference, which  present in some detail new calculations in HQET
or new results for inclusive heavy hadron decays. It is divided into two
parts, one dealing with HQET and exclusive decays and the second
devoted to inclusive decays.

In the next section we discuss the heavy mass limit of QCD and the additional
symmetries of this limit. We formulate HQET, including terms up to
order $1/m^2$. The strategy of a HQET calculation is outlined and applied
to the weak decay matrix elements relevant for the semileptonic decays
$B \to D^{(*)} \ell \nu$. In section 3 we consider the setup for
the heavy mass expansion for inclusive decays and study the rate and
differential distributions in inclusive semileptonic decays. Finally we
comment on inclusive non-leptonic processes and inclusive rare
decays and conclude.

\section{Heavy Quark Effective Theory}
\noindent
The Green functions of QCD  containing a heavy quark in general
depend on its mass $m_Q$. This mass sets a scale that is large compared to
the scale $\bar\Lambda$, which characterizes the light degrees of freedom,
$\bar\Lambda / m_Q$ is small and becomes a reasonable expansion parameter.
The leading order in this parameter corresponds to the infinite mass limit
of QCD, which corresponds to an effective theory where the degrees of freedom
related to this large scale have been removed. This effective theory, the
so-called  HQET, may be formulated as a Lagrangian field theory, and its
Lagrangian may be obtained from QCD. There are several ways to construct
this Lagrangian and the one closest to the idea of ``integrating out''
heavy degrees of freedom is discussed in \cite{MRRderivation}, where
the small components of the heavy quark spinor field are identified as
the heavy degrees of freedom and are removed by integrating over them in
the generating functional of QCD Green functions.

We shall not go through any derivation here, but rather state the result
and its relation to full QCD. We denote the heavy quark field of full QCD
by $Q$ and define
\begin{equation} \label{Qfull1}
Q_v (x) = \exp (im_Qvx) Q (x) = h_v (x) + H_v (x) ,
\end{equation}
where $v$ is a velocity ($v^2 = 1$), which is later identified with
the velocity of the heavy hadron. Extracting this phase factor from the
full QCD field $Q$ removes the dominant part $m_Q v $ of the heavy quark
momentum, since this phase redefinition corresponds to a splitting of the
heavy quark momentum according to $p = m_Q v + k$, where the residual
momentum $k$ is small, of the order of $\bar\Lambda$.
Furthermore, $h_v$ ($H_v$) is the
large (small) component field, corresponding to the projections
\begin{equation}
h_v = P_+ Q_v , \quad H_v = P_- Q_v,  \quad
P_\pm = \frac{1}{2} (1\pm \slash{v}) .
\end{equation}
The small component field $H_v$ is related to the large scale $m_Q$;
integrating out $H_v$ from the generating
functional of QCD Green functions corresponds to the replacement
\begin{equation} \label{Qfull2}
H_v = P_- \left( \frac{1}{2m + ivD} \right) i \slash{D} h_v
\end{equation}
and this yields a non-local ``Lagrangian'' of the form \cite{MRRderivation}
\begin{equation} \label{lfull}
{\cal L} = \bar{h}_v (iv D) h_v +
\bar{h}_v i \slash{D} P_- \left( \frac{1}{2m + ivD} \right) i \slash{D} h_v ,
\end{equation}
which still contains all orders in $1/m_Q$. However, the non-locality
appearing in the second term of (\ref{lfull}) may be expanded into an
infinite series of local terms, which come with increasing powers of
$1/m_Q$. Hence one may in this way establish the desired heavy mass
expansion for the Lagrangian. The first few terms of this expansion are
\begin{equation} \label{lexp}
{\cal L}  =  \bar{h}_v (iv D) h_v +
\left( \frac{1}{2m} \right) \bar{h}_v  i \slash{D} P_- i \slash{D}  h_v
 + \left( \frac{1}{2m} \right)^2 \bar{h}_v i \slash{D} P_-
(-ivD) i \slash{D}  h_v  + \cdots
\end{equation}

The non-local expression (\ref{lfull}) is still equivalent to full QCD;
in particular it is independent of the still arbitrary velocity vector
$v$. In fact, the Lagrangian (\ref{lfull}) is invariant under an
infinitesimal shift of the velocity
\begin{eqnarray} \label{repara}
&& v \to v + \delta v \qquad v \cdot \delta v = 0 \\
&& h_v \to h_v + \frac{\delta v \hskip -0.75em / \hskip 0.25em }{2}
\left(1+ P_- \frac{1}{2m + ivD} i \slash{D} \right) h_v \nonumber \\
&& iD \to - m \, \delta v \nonumber .
\end{eqnarray}
This invariance is the so-called reparametrization invariance
\cite{repara}, which has non-trivial consequences for the Lagrangian
and also for matrix elements, since it relates terms of different
orders of the $1/m$ expansion.

However, the increasing powers of $1/m_Q$ have to be compensated by the
dimension of the operators appearing in the expansion. In a field theory,
these operators are not a priori defined, since they have to be
renormalized. This renormalization leads to additional dependences on the
heavy mass, which are in general logarithmic, at least in perturbation
theory. The expansion of (\ref{lfull}) thus gives only the coefficients
of the operators at the scale $m_Q$, at which the heavy degrees of freedom
are integrated out.

The leading logarithmic corrections to the Lagrangian have been
calculated \cite{leadmass} and the result at some scale $\mu < m_Q$ is
\begin{eqnarray} \label{lexplla}
&& {\cal L}  =  C_0 \bar{h}_v (iv D) h_v   \\ \nonumber
&& + \left( \frac{1}{2m} \right) \left[
C_0 C_1 \bar{h}_v (i D )^2 h_v
- C_0 C_2 \bar{h}_v (i v D )^2 h_v
+ (-i) C_0 C_3 \bar{h}_v \sigma_{\mu \nu} iD^\mu iD^\nu h_v \right]
\\ \nonumber
&&+ \cdots
\end{eqnarray}
with the coefficients (in Feynman gauge)
\begin{eqnarray} \label{recoL}
&& C_0 = \eta ^{-8/(33-2n_f)} \quad C_1 = 1 \\
&& C_2 = 3 \eta^{-8/(33-2n_f)} - 2 \quad
C_3 = \eta^{-9/(33-2n_f)} \nonumber
\end{eqnarray}
where
$$
\eta =  \frac{\alpha_s (\mu)}{\alpha_s (m_Q)} .
$$
The fact that $C_1 = 1$
is a consequence of reparametrization invariance, which implies non-trivial
relations between the renormalizations of the various terms in the
Lagrangian; e.g. some of the renormalization constants of the second-order
terms in the Lagrangian may be calculated from the first-order ones
\cite{chen}.

\subsection{The Heavy Quark Limit and Additional Symmetries}
\noindent
The leading term of the Lagrangian (\ref{lexp}) defines the
heavy-quark limit and exhibits the two additional symmetries. The first
symmetry is the heavy-flavour symmetry, which is due to the fact that
the leading term in the heavy mass expansion is mass-independent.
If there are two heavy flavours described by the operators $b_v$
and $c_v$, then the total Lagrangian is simply the sum of the two
\begin{equation}
{\cal L}  =  \bar{b}_v (iv D) b_v  + \bar{c}_v (iv D) c_v ,
\end{equation}
which is invariant under $SU(2)_{HF}$ rotations among the two fields
\begin{equation}
\left( \begin{array}{c} b_v \\ c_v \end{array} \right) \to
U_v \left( \begin{array}{c} b_v \\ c_v \end{array} \right) \quad
U \in SU(2)_{HF} .
\end{equation}
We have put a subscript $v$ for the transformation matrix $U$, since
this symmetry only relates heavy quarks if they move with the same velocity.
In other words, there is a heavy-flavour symmetry in each velocity sector.

The second symmetry  is the heavy-quark spin symmetry. As is clear form
the Lagrangian in the heavy-mass limit, both spin degrees of freedom
of the heavy quark couple in the same way to the heavy quark; we may
rewrite the leading-order Lagrangian as
\begin{equation}
{\cal L} = \bar{h}_v^{+s} (iv D) h_v^{+s} + \bar{h}_v^{-s} (iv D) h_v^{-s},
\end{equation}
where $h_v^{\pm s}$ are the projections of the heavy quark field on a
definite spin direction $s$
\begin{equation}
h_v^{\pm s} = \frac{1}{2} (1 \pm \gamma_5 \slash{s}) h_v,
\quad s\cdot v = 0 .
\end{equation}
This Lagrangian has a symmetry under the rotations of the heavy quark
spin and hence all the heavy hadron states moving with the velocity $v$
fall into spin-symmetry doublets as $m_Q \to \infty$. In Hilbert space
this symmetry is generated by operators $S_v (\epsilon)$ as
\begin{equation}
[ h_v , S_v (\epsilon) ] = i \slash{\epsilon} \slash{v} \gamma_5 h_v
\end{equation}
where $\epsilon$ with $\epsilon^2 = -1$ is the rotation axis.
The simplest spin-symmetry doublet in the mesonic case consists of the
pseudoscalar meson $H(v)$ and the corresponding vector meson
$H^* (v,\epsilon)$, since a spin rotation yields
\begin{equation}
\exp\left(iS_v(\epsilon) \frac{\pi}{2} \right) | H (v) \rangle =
(-i) | H^* (v,\epsilon) \rangle
\end{equation}
where we have chosen an arbitrary phase to be $(-i)$. The spin-symmetry
doublets for baryons have been considered in \cite{MRRbary}, and the
general case, also valid for excited states, has been studied in
\cite{Falk}.

In the heavy-mass limit the spin symmetry partners have to be
degenerate and their splitting has to scale as $1/m_Q$. From the
Lagrangian given above, one derives the mass relation for the heavy
ground-state mesons up to terms of order $1/m_Q$
\begin{equation} \label{massrel}
M_H = m_Q + \bar\Lambda + \frac{1}{2 m_Q} \left(\lambda_1 + d_H C_3 \lambda_2
\right)
\end{equation}
where $d_H = 3$ for the $0^-$ and $d_H = -1$ for the $1^-$ meson and $C_3$ has
been given in (\ref{recoL}). Furthermore,
the parameters $\bar\Lambda$, $\lambda_1$ and $\lambda_2$ are given by
\begin{eqnarray}
&&\langle 0 | q \gamma_5 h_v | H (v) \rangle \bar{\Lambda} =
\langle 0 | q \stackrel{\longleftarrow}{ivD} \gamma_5 h_v | H (v) \rangle
\label{Lambar} \\
&& 2 M_H \lambda_1  =
\langle H (v) | \bar{h}_v  (iD)^2  h_v
              | H (v) \rangle
\label{lam1} \\
&&  6 M_H i \lambda_2  =
\langle H (v) | \bar{h}_v \sigma_{\mu \nu} iD^\mu iD^\nu h_v
              | H (v) \rangle  ,
\label{lam2}
\end{eqnarray}
where the normalization of the states is chosen to be
$\langle H (v) | \bar{h}_v h_v | H (v) \rangle = 2 M_H$. These
parameters may be interpreted as the binding energy of the heavy meson in
the infinite mass limit ($\bar\Lambda$), the expectation value of
the kinetic energy  of the heavy quark ($\lambda_1$) and its energy
due to the chromomagnetic moment of the heavy quark ($\lambda_2$)
inside the heavy meson.

All the parameters appearing in the mass relation are subject to
renormalization or suffer from ambiguities from renormalons, the latter
subject is discussed in \cite{Renormalons}. Hence quoting values for
these parameters requires a procedure
to be defined to deal with the ambiguities.

The only parameter which is easy to access is
$\lambda_2$, since it is related to the mass splitting between $H(v)$
and $H^* (v , \epsilon)$. From the $B$-meson system we obtain
\begin{equation}
\lambda_2 (m_b) = \frac{1}{4} (M_{H^*} - M_H) = 0.12 \mbox{ GeV}^2,
\end{equation}
and using the scaling (\ref{recoL}) we obtain the same value as from the
corresponding mass splitting in the charm system. This shows that indeed
the spin-symmetry partners are degenerate in the infinite mass limit and
the splitting between them scales as $1/m_Q$.

The other parameters appearing in (\ref{massrel}) are not simply
related to the hadron spectrum.
Using the pole mass for $m_Q$ in (\ref{massrel}), QCD sum rules yield for
a value of $\bar\Lambda = 570 \pm 70$ MeV \cite{reviews}. More
problematic is the parameter $\lambda_1$; from its definition one
is led to assume
$\lambda_1 < 0$; a more restrictive inequality
\begin{equation}
-\lambda_1 > 3 \lambda_2
\end{equation}
has been derived in a quantum mechanical framework in \cite{BiMotion}
and using heavy-flavour sum rules \cite{Bisumrule}.
Furthermore, there exists also a sum rule estimate \cite{BBsumrule}
for this parameter:
\begin{equation}
\lambda_1  = - 0.52 \pm 0.12 \mbox{ GeV}^2 ,
\end{equation}
which, however, has been critizised. A more extensive discussion of this
issue is given in \cite{NeubertMPL}.

In the infinite mass limit the symmetries imply relations between matrix
elements involving heavy quarks.  For a transition between
heavy ground-state mesons $H$  (either pseudoscalar or vector)
with heavy flavour $f$ ($f'$) moving with velocities $v$ ($v'$), one
obtains in the heavy-quark limit
\begin{equation} \label{WET}
\langle H^{(f')} (v') | \bar{h}^{(f')}_{v'} \Gamma h^{(f)}_v
|  H^{(f)} (v) \rangle
 = \xi (vv') \mbox{ Tr }
\left\{ \overline{{\cal H} (v)} \Gamma {\cal H}(v) \right\} ,
\end{equation}
where $\Gamma$ is some arbitrary Dirac matrix and $H(v)$ are the
representation matrices for the spin structure of the heavy mesons
\begin{equation}
{\cal H}(v) = \frac{\sqrt{M_H}}{2} \left\{ \begin{array}{l l}
       \gamma_5 P_+ & 0^- \mbox{ meson} \\
       \slash{\epsilon} P_+ & 1^- \mbox{ meson} \\
                            & \mbox{with polarization } \epsilon .
       \end{array} \right.
\end{equation}
The single form factor for these transitions,  the Isgur--Wise
function $\xi (vv')$ contains all the non-perturbative information
for the heavy-to-heavy decay. Furthermore, heavy-quark symmetry fixes
the value of $\xi$ at the point $v = v'$ to be
\begin{equation} \label{norm}
\xi (vv' = 1) = 1 ,
\end{equation}
since the current $\bar{h}^{(f')}_{v} \Gamma h^{(f)}_v$ is one of the
generators of heavy-flavour symmetry. The generalization of (\ref{WET})
to baryons may be found in \cite{MRRbary} and to excited states in
\cite{Falk}.

The symmetries also place some restrictions on the corrections which
may appear. In general, if explicit symmetry breaking is present those
form factors, which are normalized due to the symmetry, only receive
second-order symmetry-breaking corrections. This general statement is
the Ademollo--Gatto theorem \cite{AdemolloGatto}, which has been
specialized to the case of heavy-quark symmetries by Luke \cite{Luke}.

\subsection{Strategy of a HQET Calculation}
The relations (\ref{WET}) and (\ref{norm}) hold in the heavy-quark
limit, and the machinery of HQET allows us to calculate corrections to
(\ref{WET}) and (\ref{norm}).

In general there are two types of corrections. The short-distance
corrections may be calculated in perturbation theory, based on the
leading order of the $1/m_Q$ expansion of the Lagrangian. The
logarithmic ultraviolet divergences in the effective theory
correspond to logarithmic dependences on the heavy-quark mass $m_Q$
in the full theory, and renormalization group methods may be employed
to perform resummations of these logarithms. In fact, the leading
logarithmic corrections to bilinear currents are independent of the
spin structure of the current.

The second type of corrections are the power corrections of order
$1/m_Q^n$, which in
general involve long-distance physics and hence may in general not
be calculated, but have to be parametrized. As an example, consider
a matrix element of a current $\bar{q} \Gamma Q$ mediating
a transition between a heavy meson and some arbitrary state
$| A \rangle$. Using the expansion of the full QCD field (\ref{Qfull1}),
(\ref{Qfull2}) and the corresponding expansion of the Lagrangian
(\ref{lexp}), one has,  up to order $1/m_Q$:
\begin{eqnarray} \label{exp}
&& \langle A | \bar{q} \Gamma Q
              | M (v) \rangle =
\langle A | \bar{q} \Gamma h_v
              | H (v) \rangle  \\ \nonumber
&& + \frac{1}{2m_Q} \langle A | \bar{q} \Gamma
              P_- i \slash{D} h_v
              | H (v) \rangle
 -i \int d^4 x \langle A | T \{ L_1 (x) \bar{q} \Gamma h_v \}
              | H (v) \rangle
   + {\cal O} (1/m^2)
\end{eqnarray}
where $L_1$ are the first-order corrections to the Lagrangian
as given in (\ref{lexp}). Furthermore, $| M (v) \rangle $ is the
state of the heavy meson in full QCD, including all its mass
dependence, while $| H (v) \rangle$ is the corresponding state in
the infinite mass limit.

Expression (\ref{exp}) displays the generic structure of the
higher-order corrections as they appear in any HQET calculation.
There will be local contributions coming from the expansion of
the full QCD field; these may be interpreted as the corrections to
the currents. The non-local contributions, i.e.~the time-ordered
products, are the corresponding corrections to the states and thus
in the r.h.s.\ of (\ref{exp}) only the states of the infinite-mass
limit appear.

\boldmath
\subsection{An Application: $B \to D^{(*)} \ell \nu$}
\unboldmath
As an application, we shall consider the weak transition
$B \to D^{(*)} \ell \nu$, which is the most obvious,
since both the $b$ and the $c$ quark may be considered as heavy.

In general, the left-handed $B \to D^{(*)}$ transition
matrix elements are given in terms of six form factors
\begin{eqnarray}
 && \langle D (v') | \bar{c} \gamma_\mu b | B(v) \rangle   =   \sqrt{m_B m_D}
\left[ \xi_+ (y) (v_\mu + v'_\mu)
     + \xi_- (y) (v_\mu - v'_\mu) \right] \\
&& \langle D^* (v',\epsilon) | \bar{c} \gamma_\mu b | B(v) \rangle   =
       i \sqrt{m_B m_{D^*}}
\xi_V (y) \varepsilon_{\mu \alpha \beta \rho}
                       \epsilon^{*\alpha}
                       v^{\prime \beta} v^\rho  \nonumber \\ \nonumber
&& \langle D^* (v',\epsilon) | \bar{c} \gamma_\mu \gamma_5 b | B(v) \rangle
    =  \sqrt{m_B m_{D^*}} \\ \nonumber
&& \quad \left[ \xi_{A1} (y) (vv'+1) \epsilon^*_\mu
      -  \xi_{A2} (y) (\epsilon^* v)  v_\mu
      -  \xi_{A2} (y) (\epsilon^* v)  v'_\mu \right] ,
\end{eqnarray}
where we have defined $y = vv'$. In the heavy-quark limit, these
form factors are related to the Isgur--Wise function $\xi$ by
\begin{eqnarray*}
&& \xi_i (y) = \xi (y) \mbox{ for } i = +,V,A1,A3 , \\
&& \xi_i (y) = 0       \mbox{ for } i = -,A2 .
\end{eqnarray*}

The normalization statement (\ref{norm}) may be used to perform a
model-independent determination of $V_{cb}$ from semileptonic
heavy-to-heavy decays by extrapolating
the lepton spectrum to the kinematic endpoint $v=v'$.
Using the mode $B \to D^{(*)} \ell \nu$ one obtains the relation
\begin{equation} \label{extra}
\lim_{v \to v'} \frac{1}{\sqrt{(vv')^2-1}} \frac{d \Gamma}{d(vv')} =
\frac{G_F^2}{4 \pi^3} |V_{cb}|^2 (m_B - m_{D^*})^2 m_{D^*}^3
|\xi_{A1} (1)|^2 .
\end{equation}
In the heavy-quark limit the form factor $\xi_{A1}$ reduces to the
Isgur--Wise function and is unity at the non-recoil point; aside from
$|V_{cb}|$ everything in the r.h.s.\ is known.

The corrections to this relation have been calculated along the lines
outlined above in leading and subleading order. A complete discussion
may be found in the review article by Neubert \cite{reviews}, including
reference to the original papers. Here we only state the final
result
\begin{eqnarray}
\xi_{A1} (1)  &=&  x^{6/25}
\left[ 1 + 1.561 \frac{\alpha_s (m_c) - \alpha_s (m_b)}{\pi}
              - \frac{8 \alpha_s (m_c)}{3 \pi} \right.
\\
&& \quad  + z \, \left\{\frac{25}{54} - \frac{14}{27} x^{-9/25}
              + \frac{1}{18} x^{-12/25} + \frac{8}{25} \ln x
       \right\} \nonumber \\ \nonumber
&& \quad \left. - \frac{\alpha_s (\bar{m})}{\pi} \frac{z^2}{1-z} \ln z \right]
 + \, \delta_{1/m^2}  ,
\label{msquared}
\end{eqnarray}
where we use the abbreviations
$$
x = \frac{\alpha_s (m_c)}{ \alpha_s (m_b)}, \quad z = \frac{m_c}{m_b}
$$
and $\bar{m}$ is a scale somewhere between $m_b$ and $m_c$.

The contributions in the square bracket originate from leading
and subleading QCD radiative corrections. These include also
the terms of order $z^n$, which are short-distance
contributions and hence may be calculated perturbatively.

The power corrections to the normalization are summarized in the
correction terms $\delta_{1/m^2}$. The form factor $\xi_{A1}$ is
protected by Lukes theorem, i.e.~it does not receive corrections
of the order $1/m_Q$. Thus the first non-vanishing recoil corrections
are of order
$(\Lambda / m_c)^2$, $(\Lambda / m_b)^2$ and $\Lambda^2 / (m_b m_c)$.
These contributions may only be estimated, since they need an input
beyond heavy-quark effective theory. There are various estimates
for these corrections \cite{FN92}-\cite{BigiIncVcb}, which are compatible
with one another; a very recent compilation of the various results
yields \cite{NeubertVcb}
\begin{equation} \label{delm2}
\delta_{m^2} = -(5.5 \pm 2.5)\% .
\end{equation}
However, this is an estimate based on various assumptions; in fact the
estimate of $\delta_{1/m^2}$ will be the final limitation for a
model-independent extraction of $V_{cb}$ from exclusive decays.

Adding all the corrections to the normalization, the value quoted in
\cite{NeubertVcb} for the normalization is
\begin{equation}
\xi_{A1} (1) = 0.93 \pm 0.03 ,
\end{equation}
where the error of 6\% is due to the uncertainty of the $1/m^2$
corrections and the next-to-next-to-leading-order short-distance
contibutions. This leads finally to a value of $V_{cb}$; taking
into account the latest data one finds \cite{NeubertVcb}
\begin{equation}
|V_{cb}|   = 0.040 \pm 0.003 .
\end{equation}

\section{The Heavy--Mass Limit for Inclusive Decays}
Another important development in heavy-flavour physics was the
formulation of the heavy-mass expansion for inclusive decays
\cite{russnl}-\cite{FLS93}, including even non-leptonic processes
\cite{BigiNonl}. The main idea
is to apply the operator-product
expansion, making use of the fact that the heavy quark mass sets a
large scale. This expansion involves operators with increasing
dimension, the coefficients of which are proportional to the
appropriate power of $1/m_Q$. The mass dependence of the matrix
elements of these operators may as well be expanded in powers of
$1/m_Q$ using the machinery of HQET, and hence one
may set up a $1/m_Q$ expansion for inclusive rates
and also for differential distributions; generically the leading
term of this expansion is the decay of a free quark.

Applying this idea to the energy spectra of the charged lepton in
inclusive semileptonic decays of heavy mesons, the relevant expansion
parameter is not $1/m_Q$, but rather $1/(m_Q - 2E_\ell  )$; the
denominator is thus the energy release of
the decay. In almost all phase space the energy release is of
the order of the heavy mass; it is only in the endpoint region that it
becomes small and hence the expansion breaks down. This problem
may be fixed by a resummation of terms in the operator product
expansion, which strongly resembles the summation corresponding to
leading twist in deep inelastic scattering. Analogously to the
parton-distribution function, a universal function appears, which
determines all inclusive heavy-to-light decays.

\subsection{Operator Product Expansion}
The general effective Hamiltonian for a decay of a heavy
(down-type) quark
is given by
\begin{equation} \label{heff}
{\cal H}_{eff} = \bar{Q} R
\end{equation}
where the operator $R$ describes the decay products, e.g.
\begin{equation}
R_{sl} = \frac{G_F^2}{\sqrt{2}} V_{Qq} \,\, \gamma_\mu (1-\gamma_5) q
\,\, (\bar{\nu}_\ell \gamma^\mu (1-\gamma_5) \ell)
\end{equation}
for semileptonic decays.

The inclusive decay rate for a heavy hadron containing the quark $Q$
is then given by
\begin{eqnarray} \label{inclusive}
\Gamma  &\propto &  \sum_X (2 \pi)^4 \delta^4 (P_B - q - P_X )
| \langle X_s | {\cal H}_{eff} | B(v) \rangle |^2
\\ \nonumber
 &=&
\int d^4 x  \langle B(v) |{\cal H}_{eff} (x)
            {\cal H}_{eff}^\dagger (0) | B(v) \rangle
\\ \nonumber
 & =&  2 \mbox{ Im}
\int d^4 x  \langle B(v) |T \{ {\cal H}_{eff} (x)
            {\cal H}_{eff}^\dagger (0) \} | B(v) \rangle .
\end{eqnarray}
The matrix element appearing in (\ref{inclusive}) contains a
large scale, namely the mass of the heavy quark. The first step
towards a $1/m_Q$ expansion is to make this large scale explicit.
This may be  done by a phase redefinition as in (\ref{Qfull1}). This
leads to
\begin{equation}
\Gamma  \propto
2 \mbox{ Im}
\int d^4 x  e^{-im_Q vx}
 \langle B(v) |T \{ \widetilde{{\cal H}}_{eff} (x)
   \widetilde{{\cal H}}_{eff} ^\dagger (0) \} | B(v) \rangle
\end{equation}
where
\begin{equation}
\widetilde{{\cal H}}_{eff}
= \bar{Q}_v R
\end{equation}
with $Q_v$ from (\ref{Qfull1}). In this way it becomes clear
that a short-distance expansion is possible, if the mass $m_Q$ is
large.  The second step is thus to perform an operator-product
expansion, which has the general form
\begin{eqnarray*}
\int d^4 x  e^{im_Q vx}
&& \langle B(v) |T \{ \widetilde{{\cal H}}_{eff} (x)
\widetilde{{\cal H}}_{eff}^\dagger (0) \}| B(v) \rangle
\\
&& \qquad = \sum_{n=0}^\infty  \left(\frac{1}{2m_Q}\right)^n
     C_{n+3} (\mu) \langle B(v) |{\cal O}_{n+3}| B(v) \rangle_\mu ,
\end{eqnarray*}
where ${\cal O}_n$ are operators of dimension $n$, with their
matrix elements renormalized at scale
$\mu$, and $C_n$ are the corresponding Wilson coefficients.

In a third step one removes the mass dependences from the matrix elements
by expanding the heavy quark fields appearing in the operators ${\cal O}_n$
using  (\ref{Qfull1}) and (\ref{Qfull2}), as well as the states by including
the corrections to the Lagrangian as time-ordered products.

The lowest-order term of the operator product expansion is a
scalar dimension-3 operator and hence it is either
$\bar  Q_v \slash{v} Q_v$ or $\bar Q_v Q_v$. The first one
is the $Q$-number current that is normalized even in full QCD,
while the second may be related to the first via
\begin{equation}
 \bar Q_v Q_v = v_\mu \bar Q_v \gamma_\mu Q_v
 + \frac{1}{2 m_Q}
\bar{h}_v  \left[ (iD)^2 - (ivD)^2 +
\frac{i}{2} \sigma_{\mu \nu} G^{\mu \nu} \right] h_v
 + {\cal O} (1/m_Q^3) .
\end{equation}
where $G_{\mu \nu}$ is the gluon field strength.
Evaluating its contribution yields the free quark decay rate.

All dimension-4 operators are proportional to the
equation of motion ${\cal O}_4 \propto \bar Q_v (iv D) Q_v$,
and the first non-trivial contribution comes from
dimension-5 operators and are of order of $1/m_Q^2$.
For mesonic decays there are only the two parameters $\lambda_1$
and $\lambda_2$ given in (\ref{lam1}) and (\ref{lam2}),
which parametrize the non-perturbative input in the order
$1/m_Q^2$.

\subsection{Inclusive Semileptonic Decays}
Applying the method outlined in the last paragraph to
semileptonic decays, one finds for the decay
$B \to X_c \ell \nu$
\begin{equation} \label{bcsl}
\Gamma(B \to X_c \ell \nu)
 = \Gamma_b |V_{cb}|^2
\times\left[
      \left(1+\frac{\lambda_1}{2m_c^2}\right) f_1 \left(\frac{m_c}{m_b}\right)
      -\frac{9\lambda_2}{2m_c^2} f_2 \left(\frac{m_c}{m_b}\right)
      \right] ,
\end{equation}
where
\begin{equation}
\Gamma_b = \frac{G_F^2 m_b^5}{192\pi^3}
\end{equation}
and the two $f_j$ are phase-space functions given by
\begin{eqnarray} \label{f1}
f_1(x) &=& 1-8x^2+8x^6-x^8-24x^4\log x ,
\\  \nonumber
f_2(x) &=& 1-\frac{8}{3}x^2-8x^4+8x^6+\frac{5}{3}x^8+8x^4\log x  .
\end{eqnarray}
{}From this one may read off the result for
$B \to X_u \ell \nu_\ell$
\begin{equation} \label{busl}
\Gamma(B \to X_u \ell \nu) = \Gamma_b |V_{ub}|^2
 \left[
      1+\frac{\lambda_1 - 9 \lambda_2}{2m_b^2}
      \right]   .
\end{equation}
Expressions (\ref{bcsl}) and (\ref{busl}) contain the leading
non-perturbative corrections, parametrized by $\lambda_1$ and
$\lambda_2$. However, before this may be confronted with data, one has
to apply as well QCD radiative corrections, which have been studied in
detail \cite{QCDradcorr}, \cite{BallBraun}.

The method of the operator-product expansion may also  be used to
obtain the non-perturbative corrections to the charged lepton energy
spectrum. In this case the procedure outlined in the last paragraph is
applied not to the full effective Hamiltonian, but rather only to the
hadronic currents. The rate is written as a product of the hadronic and
leptonic tensor
\begin{equation}
d \Gamma = \frac{G_F^2}{4 m_B} | V_{Qq} |^2 W_{\mu \nu}
\Lambda^{\mu \nu} d(PS) ,
\end{equation}
where $d(PS)$ is the phase-space differential.
The short-distance expansion is then performed for
the two currents appearing in
the hadronic tensor. Redefining the heavy-quark fields as in
(\ref{Qfull1}) and (\ref{Qfull2}) one finds that the momentum transfer variable
relevant for the short-distance expansion is $m_Q v - q$, where
$q$ is the momentum transfer to the leptons.

The structure of the expansion for the spectrum is identical to the one
of the total rate. The contribution of the dimension-3 operators
yields the free-quark decay spectrum, there are no contributions from
dimension-4 operators, and the $1/m_b^2$ corrections are parametrized
in terms of $\lambda_1$ and $\lambda_2$. Calculating the spectrum for
$B \to X_c \ell \nu$ yields relatively complicated expressions, which may
be found in \cite{Bigiincsl}-\cite{tmincsl}. However, for the decay
$B \to X_u \ell \nu$ the expression simpifies and is given by
\begin{eqnarray} \label{buspec}
\frac{1}{\Gamma_b} \frac{d\Gamma}{dy}  &=&  2y^2 (3-2y)
        + \frac{10y^2}{3} \frac{\lambda_1}{m_b^2}
        + 2y(6+5y) \frac{\lambda_2}{m_b^2}  \nonumber \\
&&  - \frac{\lambda_1 + 33 \lambda_2}{3m_b^2} \delta (1-y)
  -  \frac{\lambda_1}{3m_b^2} \delta ' (1-y)   ,
\end{eqnarray}
where $y = 2 E_\ell / m_b $ is the rescaled energy of the charged lepton.

\begin{figure}
   \epsfysize=9cm
   \centerline{\epsffile{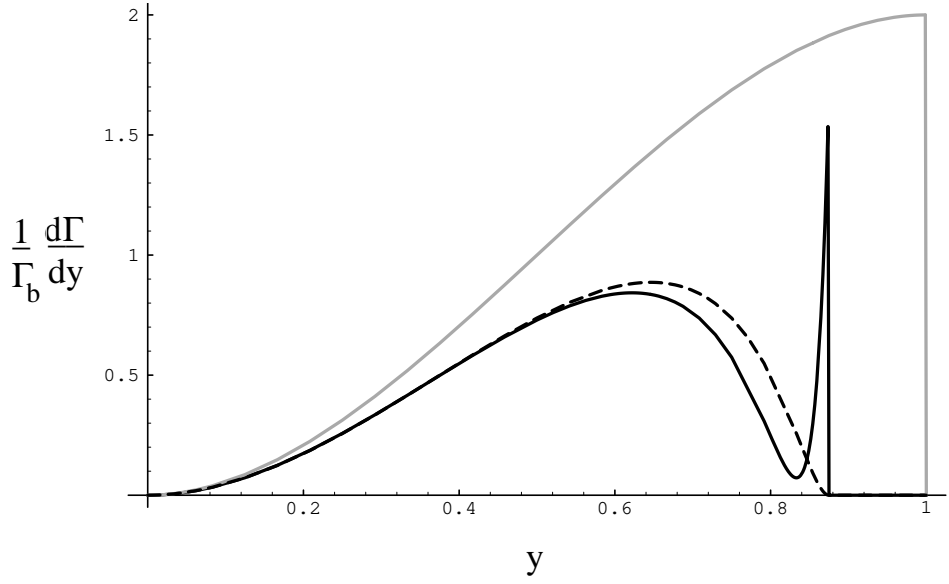}}
   \caption{The electron spectrum for free quark
$b\rightarrow c$ decay (dashed line), free quark $b\rightarrow u$ decay
(grey line), and  $B\rightarrow X_c e \bar\nu_e$ decay including
$1/m_b^2$ corrections (solid line) with $\lambda_1 = - 0.5$ GeV${}^2$
and $\lambda_2 = 0.12$ GeV${}^2$. The figure is from
\protect{\cite{mwincsl}}.}
\label{fig5}
\end{figure}

Figure~\ref{fig5} shows the distributions for inclusive semileptonic
decays of $B$ mesons. The spectrum close to the endpoint, where the
lepton energy becomes maximal, exhibits a sharp spike as $y \to y_{max}$.
Close to the endpoint we have
\begin{eqnarray}
&& \frac{1}{\Gamma_b} \frac{d\Gamma}{dy}  \sim  2 \Theta(1-y-\rho)
\\ \nonumber
&& \times \left[ \frac{\lambda_1}{(m_Q (1-y))^2}
\left(\frac{\rho}{1-\rho} \right)^2
\left\{ 3 - 4 \left(\frac{\rho}{1-\rho} \right) \right\} \right]
\end{eqnarray}
where $\rho = m_c^2 / m_b^2$. This expression behaves like
$\delta$-functions and its derivatives as $\rho \to 0$, which can be
seen in (\ref{buspec}). This behaviour indicates a breakdown of the operator
product expansion close to the endpoint, since for the spectra the
expansion parameter is not $1/m_Q$, but rather $1/(m_Q - qv)$, which
becomes $1/(m_Q [1-y])$ after the integration over the neutrino momentum.
In order to obtain a description of the endpoint region, one has to perform
some resummation of the operator product expansion.

\subsection{The Endpoint Region $E_\ell \sim E_{max}$}
Very close to the endpoint of the inclusive semileptonic decay spectra
only a few resonances contribute. In this resonance region one may
not expect to have a good description of the spectrum using
an approach based on parton-hadron duality; here a sum over a few
resonances will be appropriate.

In the variable $y$ the size of this region is however of the order of
$(\bar\Lambda / m_Q)^2$ and thus small. In a larger region of the order
$\bar\Lambda / m_Q $, which we shall call the endpoint region,
many resonances contribute and one may hope to describe the spectrum
in this region using parton-hadron duality.

It has been argued in \cite{NeubertShape} that the $\delta$-function-like
singularities appearing in (\ref{buspec}) may be reinterpreted as
the expansion of a non-perturbative function describing the spectrum in the
endpoint region. Keeping only the singular
terms of (\ref{buspec}) we write
\begin{equation} \label{shape}
\frac{1}{\Gamma_b} \frac{d\Gamma}{dy}  = 2y^2 (3-2y) S(y) ,
\end{equation}
where
\begin{equation} \label{momexp}
S(y) = \Theta (1-y) + \sum_{n=0}^{\infty} a_n \delta^{(n)} (1-y)
\end{equation}
is a non-perturbative function given in terms of the moments of the
spectrum, taken over the endpoint region. These moments themselves have an
expansion in $1/m_Q$ such that $a_n \sim 1/m_Q^{n+1}$,
and we shall consider only the leading term in
the expansion of the moments, corresponding to the most singular
contribution to the endpoint region.

Comparing (\ref{buspec}) with
(\ref{shape}) and (\ref{momexp}) one obtains that
\begin{eqnarray}
a_0 &=& \int dy (S(y) - \Theta (1-y)) = 0  \\ \label{a1}
a_1 &=& \int y  (S(y) - \Theta (1-y)) = -\frac{\lambda_1}{3 m_Q^2}
\end{eqnarray}
where the integral extends over the endpoint region.

The non-perturbative function implements a resummation of the
most singular terms contributiong to the endpoint and, in the language
of deep inelastic scattering, corresponds to the
leading twist contribution. This
resummation has been studied in QCD \cite{ManNeu,BiMotion} and the
function $S(y)$ may be related to the distribution of the light cone
component of the heavy quark residual momentum
inside the heavy meson. The latter is a fundamental
function for inclusive heavy-to-light transitions, which has been
defined in \cite{BiMotion}
\begin{equation}\label{fdef}
   f(k_+) = \frac{1}{2m_B}
   \langle B(v)|\,\bar h_v\,\delta(k_+-i D_+)\,h_v\,
   |B(v)\rangle ,
\end{equation}
where $k_+ = k_0 + k_3$ is the positive light cone component
of the residual momentum $k$. The relation between the two functions
$S$ and $f$ is given by
\begin{equation}
S(y) = \frac{1}{m_Q} \int\limits_{-m_Q (1-y)}^{\bar\Lambda} dk_+ f(k_+)
\end{equation}
from which we infer that the $n^{th}$ moment of the endpoint region
is given in terms of the matrix element $\langle B(v)|
\bar h_v (i D_+)^n h_v |B(v)\rangle $.

The function $f$ is a universal distribution function, which
appears in all heavy-to-light inclusive decays; another example
is the decay $B \to X_s \gamma$ \cite{NeubertBsg,BiMotion},
where this function determines
the photon-energy spectrum in a region of order $1/m_Q$ around
the $K^*$ peak.

\begin{figure}[t]
%   \vspace{0.5cm}
   \epsfysize=9cm
   \centerline{\epsffile{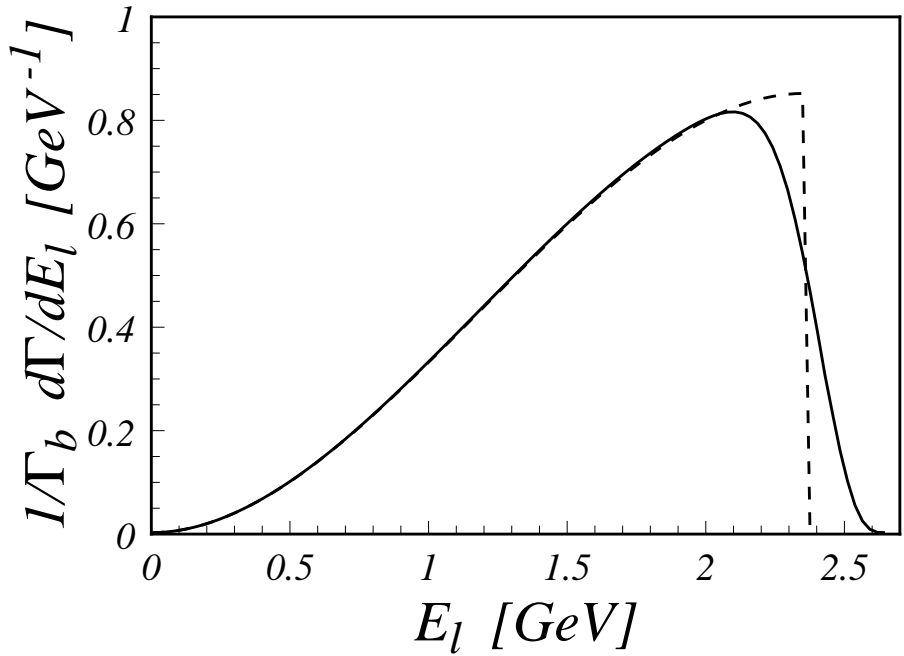}}
   \caption{Charged-lepton spectrum in $B\to X_u \ell  \bar\nu$
decays. The solid line is (\protect{\ref{shape}}) with the ansatz
(\protect{\ref{ftoy}}), the dashed
line shows the prediction of the free-quark decay model. The figure
is from \protect{\cite{ManNeu}}.}
\label{fig2}
\end{figure}

Some of the properties of $f$ are known. Its support is
$-\infty < k_+ < \bar\Lambda$, it is normalized to unity, and its
first moment vanishes. Its second moment is given by $a_1$, and its
third moment has been estimated \cite{BiMotion,Ma94}. A one-parameter
model for $f$ has been suggested in \cite{ManNeu}, which incorporates
the known features of $f$
\begin{equation} \label{ftoy}
   f(k_+)  =  {32\over\pi^2\bar\Lambda}\,(1-x)^2
    \exp\bigg\{ - {4\over\pi}\,(1-x)^2 \bigg\}\,
   \Theta(1-x)  ,
\end{equation}
where $x = k_+ / \bar\Lambda$, and
the choice $\bar\Lambda = 570$ MeV yields reasonable values
for the moments. In fig.~\ref{fig2} we show the spectrum for
$B \to X_u \ell \nu_\ell$ using the ansatz (\ref{ftoy}).

Including the non-perturbative effects yields a reasonably behaved
spectrum in the endpoint region and the $\delta$-function-like
singularities have disappeared. Furthermore, the spectrum now extends
beyond the parton model endpoint;
it is shifted from $E_\ell^{max} = m_b / 2$ to the physical
endpoint $E_\ell^{max} = M_B / 2$, since $f$ is non-vanishing for
positive values of $k_+ < \bar\Lambda = M_B - m_b$.

\subsection{Inclusive Non-leptonic and Rare Decays}
The same method as described above for the semileptonic decays
has been applied to rare \cite{FLS93} and also to non-leptonic decays,
see \cite{BigiNonl} for a recent review. For simplicity
we shall restrict the discussion here to a few simple examples.

The effective Hamiltonian for the radiative rare decays
$B \to X_s \gamma$ is given by
(\ref{heff}) with
\begin{equation}
R = - \frac{4 G_F}{\sqrt{2}} C_7(m_b) \frac{e}{32\pi^2} m_b
      \,\, \sigma^{\mu\nu} (1+\gamma_5) s \,F_{\mu\nu} ,
\end{equation}
where $C_7(\mu)$ is a coefficient calculated e.g. in \cite{C7}.

Going through the steps outlined above, one obtains for the inclusive
rate of $B \to X_s \gamma$, including the first non-perturbative
correction \cite{FLS93}
\begin{equation}
\Gamma_{B\to X_s\gamma}
     =  \frac{\alpha G_F^2}{16\pi^4}
     m_b^5 |V_{ts} V_{td}^*|^2 |C_7(m_b)|^2
     \left[1+\frac{1}{2m_b^2}
    \left(\lambda_1-9\lambda_2 \right)\right] .
\end{equation}
Along the same lines one may study the inclusive decays
$B\to X_s \ell^+ \ell^-$, the results for the lepton spectra
and the total rates may be found in \cite{FLS93}.

Similarly one may consider the total rates for inclusive
non-leptonic decays. Neglecting for simplicity penguin contributions
and CKM-suppressed decay modes, the effective Hamiltonian is given by
\begin{equation}
H_{eff} = \frac{G_F}{\sqrt{2}} \left( C_1 (\mu) O_1(\mu) +
                               C_2 (\mu) O_2 (\mu) \right)
\end{equation}
with the two operators
\begin{eqnarray}
&& O_1 = \{\bar b \gamma_\mu (1-\gamma_5) c\}
      \{\bar u \gamma_\mu (1-\gamma_5) d\} ,
\\
&& O_2 = \{\bar b \gamma_\mu (1-\gamma_5) d\}
      \{\bar u \gamma_\mu (1-\gamma_5) c\} ,
\end{eqnarray}
where the braces denote the coupling to colour singlet,
and the $C_j$ are the corresponding Wilson coefficients,
which may be found e.g.~in \cite{C12}.

{}From this one obtains for the inclusive width for $B \to X_c$
\cite{BigiNonl}
\begin{equation} \label{nonlep}
\Gamma_{B\to X_c} =
\frac{G_F^2 m_b^5}{64 \pi^3}
\left\{ A_1 f_1 \left(\frac{m_c}{m_b} \right)
\left[ 1 + \frac{1}{2 m_b^2} (\lambda_1 - 9 \lambda_2) \right]
 - 48 A_2 f_3 \left(\frac{m_c}{m_b}\right)
\frac{1}{2m_b^2} \lambda_2 \right\} ,
\end{equation}
where $A_{1/2}$ is a combination of Wilson coefficients appearing in
the effective Hamiltonian
\begin{eqnarray}
&& A_1 = C_1^2 (m_b) + C_2^2 (m_b) + \frac{2}{3} C_1 (m_b) C_2 (m_b)
\\ \nonumber
&& A_2 = \frac{2}{3} C_1 (m_b) C_2 (m_b) ,
\end{eqnarray}
and $f_1$ and $f_3$ are phase-space factors; $f_1$ is defined
in (\ref{f1}), while $f_3$ is
\begin{equation}
f_3 (x) = (1-x^2)^3 .
\end{equation}

Equation(\ref{nonlep}) is a QCD-based calculation of
the inclusive non-leptonic
width, and together with the corresponding expression
for the semileptonic width (\ref{bcsl}) it
gives us the lifetime of bottom hadrons and their semileptonic
branching fraction, including the
non-perturbative corrections in a $1/m_Q$ expansion. However, before
one may compare the results with data, one has to take into account
perturbative QCD corrections as well. These corrections have been
calculated  and are presented in a contribution to this conference
\cite{BallMPL}.

\section{Concluding Remarks}
The expansion of QCD in inverse powers of the heavy-quark mass has
put heavy-quark physics on a model-independent basis. In particular,
the symmetries present in the heavy quark limit allow a variety of
model-independent predictions for weak decay matrix elements. For
mesons, the Isgur--Wise function (\ref{WET})
is the only non-perturbative input in the heavy mass
limit.

However, the corrections of order $\bar\Lambda /m_Q$ introduce in
general new form factors, i.e.~an additional
non-perturbative input is needed. Still a few relations, like the
normalization of certain form factors, are do not receive linear
corrections, and the first subleading contribution is of order
$(\bar\Lambda /m_Q)^2$. For $m_Q = m_c = 1.5$ GeV and for
$\bar\Lambda = 500$ MeV this gives a typical size of corrections
in the ballpark of 10\%, which is what is found, for example for the
normalization of the form factor relevant for the $V_{cb}$
determination. In general, these corrections will be the final
limitation for model-independent statements from HQET.

These remarks apply in particular to the determination of $V_{cb}$
from the exclusive channel $B \to D^* \ell \nu_\ell$, where the
theoretical errors quoted above is about 6\% and dominated
by the uncertainties of
the estimates of the $1/m_Q^2$ contributions, which need model input.
This has to be compared with the determination of $V_{cb}$
from inclusive decays, which has been discussed in detail in
\cite{BigiIncVcb}. The inclusive width in the framework discussed above
depends on the quark masses, and superficially one finds a $m_Q^5$
dependence. This would mean that even small uncertainties in the
heavy-quark mass would have a large effect on a $V_{cb}$ determination
based on
the inclusive width. However, it has been argued in \cite{BigiIncVcb}
that the inclusive decays $B \to X_c \ell \nu$ will receive their
major contribution from the kinematic region close to the non-recoil
point; in this region the inclusive width depends almost linearly
only on the mass difference $m_b - m_c$ with only a weak dependence on
$m_b + m_c$. The quark-mass difference is much better known than the
individual masses; using the mass formula (\ref{massrel})
only $\lambda_1$ enters. Based on
these observations it has been argued in \cite{BigiIncVcb} that a
determination of $V_{cb}$ from inclusive decays may have a theoretical
error of as low as 5\%; the uncertainty here enters through the parameter
$\lambda_1$, which is at present only poorly known, but may be measured
in the future from the inclusive semileptonic decay spectra
\cite{BigiKin}. Given the present situation both methods have comparable
theoretical uncertainties and it remains to be seen for which of the
two methods the uncertainties appearing at order $1/m_Q^2$ will be
better under control.

Although there has been some theoretical progress in setting up a
QCD-based calculation for inclusive widths, non-leptonic decays still remain
a problem. It has been noticed soon after the formulation of the $1/m_Q$
expansion for inclusive non-leptonic processes
that the non-perturbative effects calculated in this way are
small, too small to explain the experimental data on the inclusive
semileptonic branching fraction of $B$ mesons. However, there are
perturbative corrections as well, which have been calculated
recently, taking into account a non-zero mass for the quarks in the final
state \cite{BallBraun,BallMPL}. These corrections are substantial only in the
channel $b \to \bar{c} c s$ and hence yield an enhancement charm
production in $B$ decays that is not supported by present data.
Thus the problem of the semileptonic branching fraction still persists.

The difficulty seems to be the calculation of the inclusive non-leptonic width,
and not the semileptonic one. This is supported by another problem, which
is the lifetime of the $\Lambda_b$ baryon. Based on the $1/m_Q$ expansion
one would conclude that the $\Lambda_b$ lifetime should be slighly smaller
than the $B$ meson lifetime, $\tau_{\Lambda_b} \sim 0.9 \tau_B$
\cite{BigiNonl}. This is not supported by recent data, indicating
that  $\tau_{\Lambda_b} \sim 0.7 \tau_B$ where the experimental error is
15\% \cite{ALEPHLambda}. The situation in the charm system is even worse,
here the lifetime differences are substantial,
$\tau_{\Lambda_c} \sim 0.5 \tau_{D^0}$ and
$\tau_{\Lambda_c} \sim 0.2 \tau_{D^\pm}$.
This indicates that the $1/m_Q$ expansion for
inclusive non-leptonic decays is not yet understood and the problems
have been recently summarized in \cite{inconclusive}. Unlike
exclusive non-leptonic decays, which still may be described
only in a model framework, the description of inclusive non-leptonic
decays is based on QCD and the above problems certainly deserve
further study.

\end{document}